\documentclass[aps,prl,amsmath,floats,floatfix, twocolumn,
superscriptaddress,nofootinbib,showpacs]{revtex4-1}

\usepackage[colorlinks, citecolor=blue, pdfborder={0 0 0}, plainpages=false]{hyperref}
\usepackage{graphicx}
\usepackage{xspace}
\usepackage[usenames,dvipsnames]{color}
\usepackage{amssymb}

\usepackage{ulem}
\newcommand{\Caltech}{\affiliation{
    TAPIR,
    Walter Burke Institute for Theoretical Physics,
    California Institute of Technology, Pasadena, CA 91125, USA}}
\newcommand{\Cornell}{\affiliation{Center for Radiophysics and Space
    Research, Cornell University, Ithaca, New York 14853, USA}}
\newcommand{\CITA}{\affiliation{Canadian Institute for Theoretical
    Astrophysics, 60 St.~George Street, University of Toronto,
    Toronto, ON M5S 3H8, Canada}} %
\newcommand{\AEI}{\affiliation{Albert Einstein Institute, Max-Planck-Institut f\"ur Gravitationsphysik, 
    Potsdam, Germany}} %
\newcommand{\JPL}{\affiliation{Jet Propulsion Laboratory, California Institute of Technology, Pasadena, 
    CA 91106, USA}}
\newcommand{\SYR}{\affiliation{Department of Physics, Syracuse University, Syracuse, NY 13244, USA}}
\newcommand{\Berk}{\affiliation{Lawrence Berkeley National Laboratory, Berkeley, CA 94720, USA}}
\newcommand{\WashSt}{\affiliation{Department of Physics and Astronomy, Washington State University, Pullman, Washington 99164, USA}}
\newcommand{\difest}{\affiliation{DiFeST, University of Parma, and INFN Parma I-43124 Parma, Italy}}

\begin{document}

\title{
Gravitational waveforms for neutron star
binaries from binary black hole simulations}

\author{Kevin Barkett}\thanks{kbarkett@caltech.edu} \Caltech
\author{Mark A. Scheel} \Caltech
\author{Roland Haas} \AEI \Caltech
\author{Christian~D.~Ott} \Caltech
\author{Sebastiano Bernuzzi} \Caltech \difest
\author{Duncan A. Brown} \SYR
\author{B\'ela Szil\'agyi} \JPL
\author{Jeffrey~D.~Kaplan} \Caltech
\author{Jonas Lippuner} \Caltech
\author{Curran D.\ Muhlberger} \Cornell
\author{Francois Foucart} \Berk\CITA
\author{Matthew D. Duez} \WashSt

\date{\today}

\begin{abstract}

Gravitational waves from binary neutron star (BNS) and black hole/neutron
star (BHNS) inspirals are primary sources
for detection by the Advanced Laser Interferometer Gravitational-Wave
Observatory. The tidal forces acting on the 
neutron stars induce changes in
the phase evolution of the gravitational waveform, and these changes can be used
to constrain the nuclear equation of state. Current methods of generating BNS
and BHNS waveforms rely on either computationally challenging full 3D hydrodynamical 
simulations or approximate analytic solutions. We introduce a new method for
computing inspiral waveforms for BNS/BHNS systems by adding the
post-Newtonian (PN) tidal effects to full numerical simulations of binary black
holes (BBHs), effectively
replacing the nontidal terms in the PN expansion with BBH results. Comparing a
waveform generated with this method against 
a full hydrodynamical simulation of a BNS inspiral yields a phase
difference of $<1$ radian over $\sim 15$ orbits.
The numerical phase accuracy required of BNS simulations to measure the accuracy
of the method we present here is estimated as a function of the tidal
deformability parameter $\lambda$.

\end{abstract}


\maketitle


\newcommand{\vb}{\bar{v}}
\newcommand{\tb}{\bar{t}}

\section{I. Introduction}
In September 2015, the Advanced Laser Interferometer Gravitational-Wave
Observatory(aLIGO) directly detected, for the first time ever, gravitational
waves (GWs)~\cite{LIGOVirgo2016a} and the network of observatories will be
joined shortly by advanced Virgo~\cite{aVirgo2} and KAGRA~\cite{kagra}. The most
likely GW sources for these detectors are mergers of binaries consisting of
neutron stars (NSs) or black holes (BHs)~\cite{Abadie:2010cfa}. If both objects
in the binary are NSs (BNS), or if one is a NS and the other is a BH (a BHNS
binary), then the tidal deformability of the NS will alter the GW signal in a
way that is dependent upon the NS equation of state (EOS), allowing these
observatories to constrain the EOS~\cite{Flanagan2008, Hinderer2010, damour:12,
DelPozzo:13, Maselli:2013rza, ReadEtAl2013, Wade:2014vqa, Lackey2014,
Agathos:2015a}. It is therefore of key importance to understand and model the
influence of tidal effects on BNS and BHNS waveforms. We show here that a binary
black hole (BBH) waveform can be augmented with PN tidal effects to accurately
model a BNS system during the inspiral portion of the binary evolution. In
principle, this method should also be applicable to BHNS systems.

BNS waveforms are typically computed using post-Newtonian (PN) methods, which
are perturbative expansions in
the invariant velocity $v=(M d\phi/dt)^{1/3}$, where $M$ is the total mass of the system and $\phi$ is the orbital phase (here we assume $c=G=1$).
For binaries consisting of nonspinning point particles, the expansion is known
through 3.5PN order~\cite{Blanchet05}.
The static NS tidal effects first enter at 5PN order
and depend upon the tidal deformability $\lambda_i$~\cite{Vines2011}.
The parameter $\lambda_i$ measures how much each NS $i$ deforms in the presence
of a tidal field, and depends on the NS mass and EOS implicitly through its
dimensionless Love number $k_{2,i}$ and radius $R_i$:
$\lambda_i=(2/3)k_{2,i}R_i^5$~\cite{Flanagan2008}.
As $v$ increases throughout the inspiral, the missing 4PN,
 4.5PN, and 5PN point-particle terms can result in the late portion
of the PN waveform becoming inaccurate before the static
tidal terms are large enough to contribute. For estimating the NS tidal
deformability by using PN waveforms, the error introduced by neglecting the
higher order PN terms can be as large as the statistical errors due to noise in
the measured signals~\cite{Favata:2013rwa,Yagi:2014,Wade:2014vqa,Lackey2014}.


Effective-one-body (EOB) models that include tidal
effects~\cite{Damour:2009wj, Bini:2012gu,Bernuzzi:2014owa} 
also include the merger, and
provide better accuracy than PN by tuning higher-order
vacuum terms to numerical relativity (NR) BBH waveforms.  Although EOB
has accurately reproduced waveforms from NR
BNS simulations~\cite{Bernuzzi:2014owa,Hotokezaka:2015xka}, 
here we discuss a new and
different approach that holds considerable promise for modeling tidal
interactions during the inspiral.

The most accurate method of computing waveforms is carrying out full NR
simulations for BNS and BHNS binaries; see ~\cite{Hotokezaka:2015xka,
Bernuzzi:2014owa, ReadEtAl2013, Radice:2013cba, hotokezaka:12, Bernuzzi2012b,
Bernuzzi:2011aq, Baiotti2011, Baiotti:2010xh, Kawaguchi:2015, PannaraleEtAl2013,
Foucart:2013psa}
for recent work.
However, BNS and BHNS simulations are computationally challenging,
since they require solving not only the full Einstein equations but also
relativistic hydrodynamics with a realistic EOS. 
It is unfeasible to use NR hydrodynamic simulations alone to cover the
parameter space given the wide range of theoretically possible EOS and NS
masses.
In contrast, BBH systems are easier to simulate
with higher accuracy. Several large catalogs of BBH simulations and
resulting waveforms have been compiled~\cite{Aylott:2009ya, Ajith:2012az,
Pekowsky:2013ska, Hinder:2013oqa, SXSCatalog, Clark:2014fva, Healy:2014eua}.
\begin{figure*}
\includegraphics[width=1.01\textwidth]{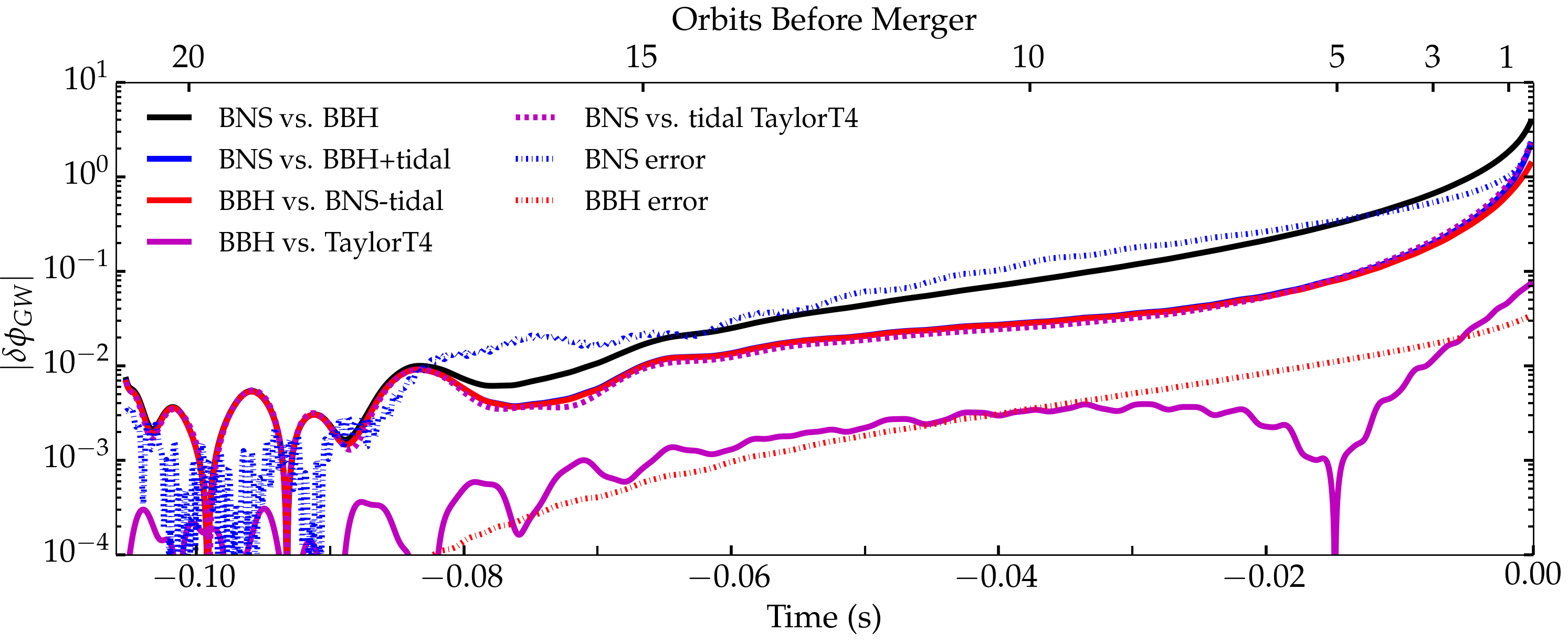}
\caption{
  Phase difference between gravitational waveforms as a function of time, for an
  equal-mass binary of nonspinning compact objects.
  Differences are shown between BNS and BBH waveforms
  (black), between a BBH waveform with TaylorT4 tidal terms added and a
  BNS waveform (blue), and between a BNS waveform with TaylorT4 tidal
  terms subtracted
  and a BBH waveform (red). The red and blue curves nearly coincide. Also shown
  are the phase differences between BBH and point-particle TaylorT4 waveforms
  (solid magenta) and between BNS and tidal TaylorT4 waveforms (dashed magenta). The numerical error in the
  BBH waveform (dashed red) and an estimate of the error in the BNS
  waveform (dashed blue) are also shown.  
  All waveforms are aligned with the BNS waveform according
    to~\cite{Boyle:2008}; the alignment time window encompasses a $5\%$ 
    change around a GW
    frequency of $280\,\rm{Hz}$ for a total mass of $M=2\times1.64M_\odot$.
  The blue and red curves are
  smaller than the black curve by a factor of $\sim 3$, demonstrating
  that tidal splicing can generate a BNS waveform from a BBH waveform 
  and vice versa.  The large error in the BNS waveform prevents us from
  fully measuring the accuracy of tidal splicing.
  \label{fig:NSNS_BBH_phase_diff}
}
\end{figure*}

We introduce here a method we call
``PN tidal splicing'', which generates BNS inspiral waveforms from NR BBH waveforms by
adding tidal interactions derived in the PN formalism, 
effectively replacing the point-particle PN terms by the numerical BBH
evolution.


We compare PN tidal splicing to NR using 
two simulations generated by SpEC~\cite{SpECwebsite}, a code developed
to evolve Einstein's equations 
and general relativistic 
hydrodynamics~\cite{Duez:2008rb,Foucart:2013a}.
The first is a new equal-mass BNS simulation with $22$ orbits before
merger~\cite{Haas:2014a}, and the neutron stars were initialized with gravitational masses $m_i\approx1.64M_\odot$ and
a polytropic EOS with $P=123.6M_{\odot}^2\rho^2$, leading to a tidal deformability of
$\lambda_i\approx5.7\times10^{36}\rm{g\ cm^2\ s^2}$.
The other is an equal-mass, nonspinning
BBH simulation~\cite{Blackman:2015pia} tagged
SXS:BBH:0180 in the public simulation catalog of the Simulating eXtreme
Spacetimes Collaboration~\cite{SXSCatalog}.
Using tidal splicing, we add tidal terms to the BBH waveform 
in an attempt to reproduce the BNS waveform. As a test, we
also subtract tidal terms from the BNS waveform in an attempt to
reproduce the BBH waveform.

Figure~\ref{fig:NSNS_BBH_phase_diff} shows that
the GW phase difference, $|\delta\phi_{\rm GW}|$, between the
`BBH+tidal' waveform and the BNS waveform is the same as 
the difference between
the `BNS$-$tidal' waveform and the BBH waveform, and both are 
a factor of $\sim 3$ smaller than the
difference between the BNS and BBH waveforms throughout the inspiral.
Thus we can mimic the inspiral
of a full BNS simulation to within a few
tenths of a radian at a fraction of the cost.
For the BBH waveform, the phase error is estimated
by the phase difference between the highest two resolutions.
The BNS simulation is a combination of spectral and finite-volume
methods with complicated convergence properties;
it is unclear how to construct an accurate error measure~\cite{Haas:2014a}.
We choose the simple
prescription of plotting the phase difference between the highest two
resolutions as a crude
error estimate. While the BBH error
estimate is small,
the error estimate in the BNS simulation is 
as large as the tidal effects themselves. Therefore,
we cannot yet fully verify the accuracy of tidal splicing until more
accurate BNS simulations are available. Below (cf. Fig.~\ref{fig:PhaseErrorVsLambda}) we will estimate the phase
accuracy required of future BNS simulations for such verification.

\section{II. Methods}
For
nonprecessing binaries, the PN equations 
for quasicircular orbits
read
\begin{align}
\frac{dv}{dt}   =&F(v)\,, \label{eq:PNeq1} \\
\frac{d\phi}{dt}=&v^3/M\,, \label{eq:PNeq2}
\end{align}
where $F(v)$
is the ratio of two functions, each known to finite PN order in $v$, 
and also depends on 
the binary's intrinsic parameters~\cite{Blanchet2014}.
Different ways of evaluating these equations result in different PN
approximants that agree to the same PN order in $v$, but diverge at higher orders.
We present methods for tidal splicing using two different approximants.


{\it TaylorT4.}
If $F(v)$ is expanded as a series in $v$ and then truncated to the
appropriate PN order, then the solution is known as the TaylorT4
approximant~\cite{Boyle2007}.
For TaylorT4, the tidal effects manifest as additional terms in the power
series for $F(v)$.
Equation~(\ref{eq:PNeq1}) can be written
\begin{align}
\frac{dv}{dt}=F(v)=F_{\rm pp}(v)+F_{\rm tid}(v)\,, \label{eq:PNeq1_tidal}
\end{align}
where $F_{\rm pp}(v)$ are the point-particle terms, and
where the additional static
tidal terms $F_{\rm tid}(v)$
are known to 6PN
order~\cite{Vines2011}.

For inspiraling PN BBHs,
$F(v)$ is governed by the point-particle terms.
PN tidal splicing uses
$\phi(t)$ from a BBH simulation together with Eqs.~(\ref{eq:PNeq1_tidal})
and~(\ref{eq:PNeq2}) [with $F_{\rm tid}(v)$ set to zero] 
to compute an accurate version of $F_{\rm pp}(v)$,
which we will call $F_{\rm NR}(v)$.
To do this, we set $\phi(t)=\phi_{\rm GW}/2$, where $\phi_{\rm GW}$ is
the GW phase of the $\ell=m=2$ spherical-harmonic
mode of the NR waveform.
Then Eq.~(\ref{eq:PNeq2}) yields 
\begin{align}
v(t)=\left(\frac{M}{2}\frac{d\phi_{\rm GW}}{dt}\right)^{1/3}\,.
\label{eq:VfromPhi}
\end{align}
Given $v(t)$, we compute $F_{\rm NR}(v)=dv/dt$ using finite differencing.
Assuming $v(t)$ is monotonic, we can write $F_{\rm NR}(v)$ as a 
single-valued function of $v$.

Using this $F_{\rm NR}(v)$ in place of $F_{\rm pp}(v)$ in 
Eq.~(\ref{eq:PNeq1_tidal}), we then re-solve Eqs.~(\ref{eq:PNeq1_tidal})
and~(\ref{eq:PNeq2}), including the tidal terms $F_{\rm tid}(v)$, to
generate a waveform for a binary that includes tidal interactions.
We express the orbital evolution of the new binary in terms of
a new time coordinate $\tb$. From 
the analytic expression for $F_{\rm tid}(v)$~\cite{Vines2011}
and Eq.~(\ref{eq:PNeq1_tidal})
we write a differential equation for $\tb$:
\begin{align}
\frac{d\tb}{dv}=\frac{1}{F_{\rm NR}(v)+F_{\rm tid}(v)}\,. \label{eq:diffeq_splicing}
\end{align}
Integrating this expression and inverting yields the function
$v(\tb)$ corresponding to the spliced waveform.

The phase of the spliced waveform,
$\bar\phi_{\rm GW}(\tb)$, is computed by integrating Eq.~(\ref{eq:PNeq2}):
\begin{align}
\bar\phi_{\rm GW}(\tb) = \frac{2}{M}\int^{\tb}_{\tb_{\rm min}}v(\tb)^3d\tb\,,
\end{align}
where $\bar t_{min}$ is the start of the numerical waveform.

In TaylorT4, the amplitude of the waveform is a function of
$v$ only, with no explicit time dependence~\cite{BFIS}. 
So here we assume
that the amplitude of the original NR waveform $A_{\rm NR}(t)$ is likewise 
a function of $v$ only, so that we can write $A_{\rm NR}(v)=A_{\rm NR}(t(v))$.  We then
use $v(\tb)$ to express this amplitude in terms of $\tb$.
In other words, the amplitude of the resulting waveform
is $\bar{A}(\tb)=A_{\rm NR}(t(v(\tb)))$. We generate a BBH waveform 
from a BNS waveform by the same method, except
we subtract instead of add
$F_{\rm tid}(v)$ in the denominator
of Eq.~(\ref{eq:diffeq_splicing}).

We require $v(t)$ to be monotonic so that $F(v)$
is single valued.
To remove high-frequency numerical noise,
the derivative in Eq.~(\ref{eq:VfromPhi}) is computed with a third order Savitzky-Golay
filter~\cite{numrec_cpp} with a window size of $\approx48.5\,\rm{\mu s}$.
This is sufficient when adding tidal
terms to the BBH
waveform considered here. However, when testing our method by subtracting
tidal terms from a BNS waveform, the phase of the BNS waveform considered
here~\cite{Haas:2014a} has large enough oscillations in $v(t)$ that
stronger smoothing is needed. We proceed by first subtracting the phase of the
TaylorT4 waveform from that of the BNS waveform, expanding this difference in
Chebyshev polynomials, truncating the Chebyshev expansion to $n=35$, and adding
back the phase of the TaylorT4 waveform. We find that the difference between the
smoothed and unsmoothed phase of the BNS waveform is less than $3\times
10^{-3}$ radians.

As discussed above, Figure~\ref{fig:NSNS_BBH_phase_diff} displays the
phase differences between NR and tidally spliced TaylorT4 waveforms.
We now examine how well \emph{pure} PN waveforms agree with NR waveforms. The
magenta solid and dashed curves in
Fig.~\ref{fig:NSNS_BBH_phase_diff} show phase differences between
TaylorT4 and BNS or BBH
waveforms. The point-particle TaylorT4 waveform does an excellent job of
reproducing the phase evolution of the BBH waveform, about at the level
of the BBH numerical error.
However, while TaylorT4 is surprisingly
accurate in the inspiral for equal-mass, nonspinning
systems~\cite{Boyle2007,MacDonald:2011ne}, this does not hold true in
general~\cite{Hannam2007c,Hannam:2010ec,Szilagyi:2015rwa}. 
Tidal splicing should be applicable to an
arbitrary BNS/BHNS system with spins and/or unequal masses, 
where there may not be an accurate PN approximant. 
References~\cite{Wade:2014vqa,Lackey2014} showed that
uncertainties in the PN waveforms are one of the largest sources of error for
tidal parameter estimation, and conclude that more accurate waveforms are
needed.


{\it TaylorF2.}
If Eqs.~(\ref{eq:PNeq1}) and~(\ref{eq:PNeq2}) are
instead converted to the frequency domain (FD) using the stationary phase
approximation before expanding the series, the
approximant is called TaylorF2~\cite{Damour:2000zb}.
TaylorF2 waveforms are expressed in the FD, and can be written
\begin{align}
\tilde{h}(f)=\tilde{A}(f)\exp\left(i\tilde\Psi(f)\right)\,,\label{eq:HFTdecomp}
\end{align}
where $\tilde{A}(f)$ is real and
$\tilde\Psi(f)$ is the Fourier phase as a function of the GW
frequency $f=v^3/(\pi M)$. 
For point particles, 
$\tilde\Psi(f)=\tilde\Psi_{\rm pp}(f)$ is known for
nonspinning systems to 3.5PN order~\cite{Damour:2000zb,Damour02}.
For tidally deformable objects, we write
$\tilde\Psi(f)=\tilde\Psi_{\rm pp}(f)+\tilde\Psi_{\rm tid}(f)$, 
where
$\tilde\Psi_{\rm tid}(f)$ has been 
calculated up to 7.5PN order, with the exception
of a few unknown constants~\cite{Bini:2012gu,damour:12}. Here we include
both 6PN tidal effects and 7.5PN tidal effects, setting the unknown
constants to 0 as was done in~\cite{Agathos:2015a}.

To add the static tidal terms to an existing BBH waveform, first the Fourier transform of
the waveform $\tilde h_{\rm NR}(f)$ is computed. The early portion of the
waveform is windowed using a Planck taper~\cite{McKechan:2010kp}
while the merger and ringdown
provide a natural windowing for the late portion. 
We then compute $\tilde\Psi_{\rm NR}(f)$ and
$\tilde A_{\rm NR}(f)$ by decomposing according to
Eq.~(\ref{eq:HFTdecomp}).
The spliced Fourier phase is then
$\tilde\Psi(f)=\tilde\Psi_{\rm NR}(f)+\tilde\Psi_{\rm tid}(f)$. Because the 
known tidal terms
do not affect the amplitude $\tilde A_{\rm NR}(f)$, the new waveform is then
\begin{align}
\tilde{h}(f)=\tilde{A}_{\rm NR}(f)\exp\left(i\left[\tilde\Psi_{\rm NR}(f)+\tilde\Psi_{\rm tid}(f)\right]\right)\,.\label{eq:htildeOfF}
\end{align}
No smoothing of the
numerical waveforms is needed for TaylorF2 splicing, unlike the TaylorT4
case.

\begin{figure}
\centering
\includegraphics[width=1.005\columnwidth]{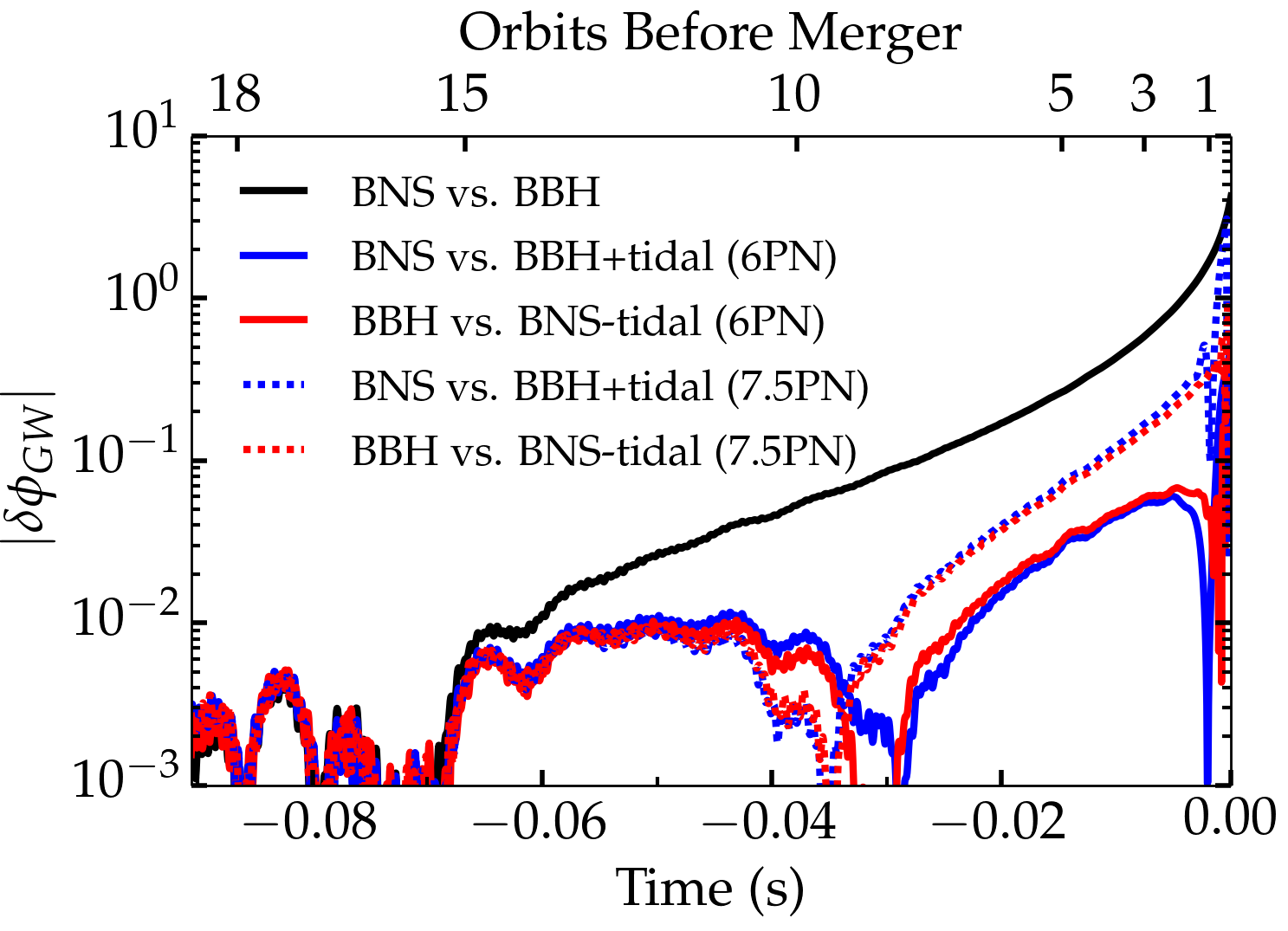}
\caption{
  The phase difference $|\delta\phi_{\rm GW}(t)|$ as a function of time for waveforms spliced with TaylorF2.
  Differences are shown between a BNS and a BBH waveform (black), between a
  BBH+tidal and a BNS
  waveform (blue), and between a BNS$-$tidal waveform
  and a BBH waveform (red) at the
  6PN (solid) and 7.5PN (dot-dashed) orders. Only the time after the windowing
  function is shown here, resulting in a shorter time axis here than in
  Fig.~\ref{fig:NSNS_BBH_phase_diff}. The late-time noise 
  is an artifact
  caused by inverse Fourier transforming 
  the unphysical high-frequency behavior of $\tilde\Psi_{\rm tid}(f)$. 
  At both PN orders, tidal splicing can generate a BNS waveform
  from a BBH waveform and vice versa.
}
\label{fig:NSNS_BBH_F2phase_diff}
\end{figure}

Since the PN approximation breaks down for high
frequencies, we impose a high frequency cutoff which we choose
to be $f_{\rm ISCO}=1/(6^{3/2}\pi M)=1338\,{\rm Hz}$, the GW
frequency corresponding to the innermost stable circular
orbit of a Schwarzschild black hole of mass equal to the
total mass of the system. It has been shown that for BNS systems, 
$f_{\rm ISCO}$ is approximately the merger frequency~\cite{Bernuzzi:2014kca}.
The starting frequency of the
NR BNS waveform after windowing is $\sim285$\,Hz.

We estimate the error of the spliced waveforms by analyzing the phase differences in the time domain after taking the inverse Fourier transform. To avoid jump
discontinuities in the Fourier phase, we roll off the effect of
$\tilde\Psi_{\rm tid}(f)$
from $f_{\rm ISCO}$ to $2\times f_{\rm ISCO}$ with a cosine window. While this
will contaminate the higher frequency content, this should allow the lower
frequencies of the inspiral to be mostly unaffected. After the inverse Fourier
transform, the time domain waveforms are
aligned in a 10\% region around $300\,$Hz. The phase
differences are shown in Fig.~\ref{fig:NSNS_BBH_F2phase_diff} and are similar to
Fig.~\ref{fig:NSNS_BBH_phase_diff}. With the exception of the last
$\sim 3\,\rm{ms}$ of the waveforms, which are affected by the high frequency
contamination, all of the spliced waveforms maintain phase differences under $0.1$ radians during most of the
inspiral, below the difference between the BNS and BBH waveforms. 
It is not clear why the 6PN terms approximate the tidal effects better 
than the 7.5PN terms; it may be because we zero the unknown constants
in the 7.5PN expression.



\begin{figure}
\centering
\includegraphics[width=1.01\columnwidth]{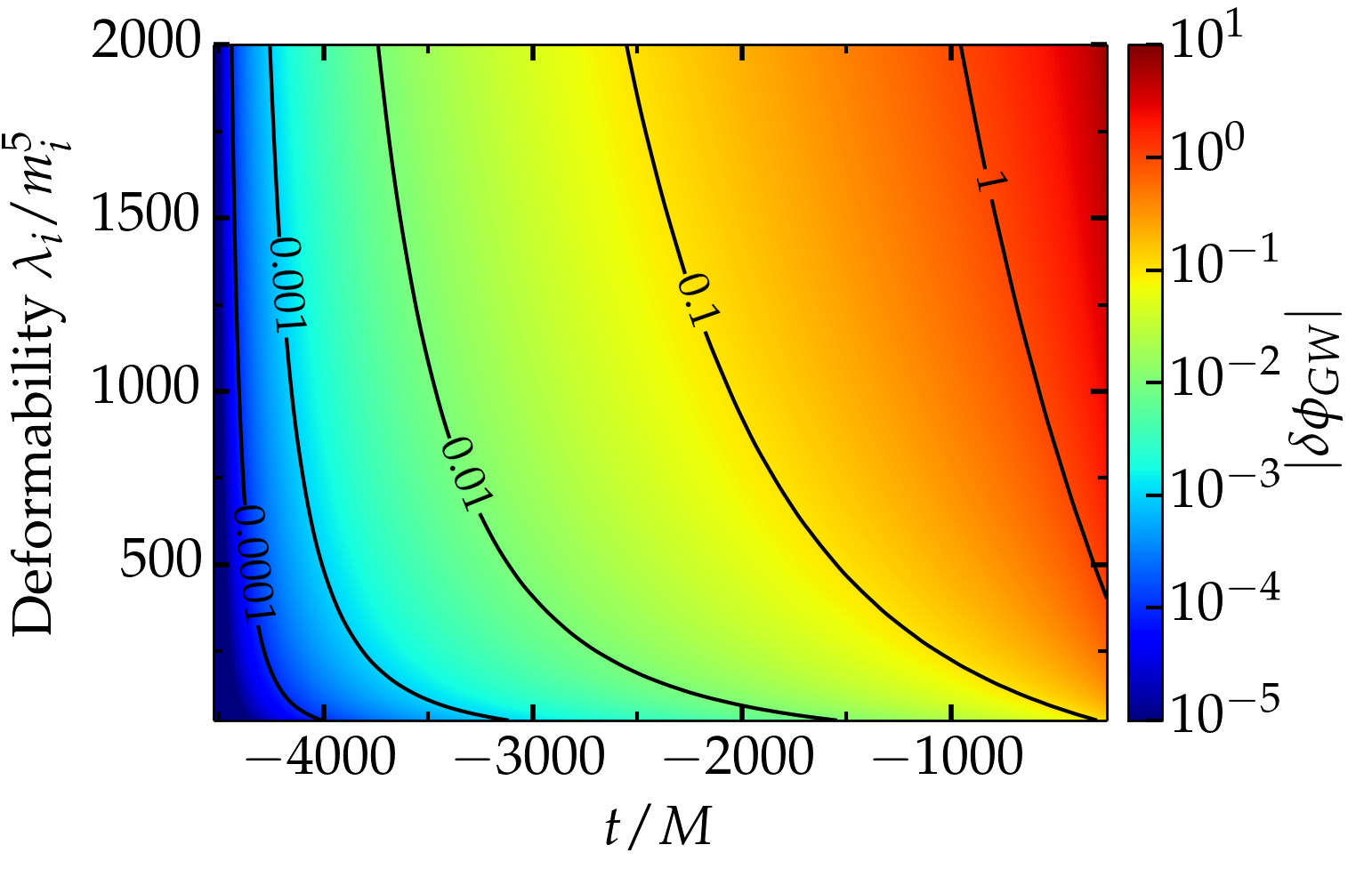}
\caption{
  Phase difference between equal-mass, 
  nonspinning BBH and `BBH+tidal' waveforms. Each
  horizontal slice through this plot shows the phase difference as a function of
  time for a particular dimensionless deformability
  $\lambda_i/m_i^5$. For our BNS simulation, $\lambda_i/m_i^5\approx453$. 
  Contours show selected values
  of the phase difference. A BNS simulation starting at dimensionless time $t/M\approx-4500$
  would need phase errors smaller than the values shown here in order to 
  measure
  tidal effects.
  Even more accurate BNS simulations would be needed to measure
  the accuracy of the tidal splicing method.
}
\label{fig:PhaseErrorVsLambda}
\end{figure}

\section{III. Discussion}
We have shown
that PN tidal splicing of
BBH waveforms can produce inspiral
waveforms for nonspinning BNS systems. This method should easily generalize
to objects with spins and to BHNS systems. 
Once a BBH
waveform is generated for a particular mass ratio and spin
configuration, it should be easy to produce BNS/BHNS waveforms via
PN tidal splicing for any EOS simply by adjusting the tidal
parameters $\lambda_i$, 
allowing the entire tidal parameter space for inspiral
waveforms to be spanned.

The accuracy of this method is limited by that of the PN tidal
terms. In particular, additional finite size effects not captured
by the current static tidal PN terms can influence waveform amplitude and phase,
and dynamical tidal effects can also contribute to the phase
evolution~\cite{Hinderer:2016a}, especially as the NSs approach merger.
 This method in principle can be improved with better PN tidal terms. Unfortunately, it is currently difficult to fully measure
the accuracy of tidal splicing until higher-accuracy many-orbit BNS simulations
are available for multiple masses and EOS.

Figure~\ref{fig:PhaseErrorVsLambda} estimates the accuracy needed for equal
mass, nonspinning BNS simulations
to see the tidal effects on the
inspiral phase of the waveform. Even smaller BNS errors would be necessary to constrain
the accuracy of tidal splicing.
We chose the start time in Fig.~\ref{fig:PhaseErrorVsLambda}
so that the inspiral spans a large enough
frequency range for aLIGO to recover 97\% of the information about
$\lambda_i$, according to the analysis presented in Fig.~3
of~\cite{damour:12}.
We assume $M=2.8M_{\odot}$ (corresponding to a prototypical NS mass of $1.4M_\odot$) and an upper frequency cutoff of
$f_{\rm{ISCO}}$.

An alternative to computing tidal terms to a higher PN order is
  to resum them in some way, as is done in~\cite{Bini:2014, Bernuzzi:2014owa,
  Hotokezaka:2015xka} in the context of EOB. It is not clear how to do this with
  tidal splicing.

Additionally, the merger/ringdown cannot be
modeled with splicing alone, especially 
for BNS mergers and BHNS systems that undergo tidal disruption.
One possibility is to combine 
an analytic waveform in the very early inspiral with a spliced
BBH waveform in the mid to late inspiral and then with a waveform from a full hydrodynamical
simulation for the merger and ringdown, to create a ``tribridized'' waveform.
This might reduce the need
for expensive hydrodynamical simulations lasting many orbits.
If necessary,
surrogate models~\cite{Field:2013cfa,Blackman:2015pia,Purrer:2014} that cover the parameter space including the EOS can be forged
from spliced BBH waveforms. 

\section{Acknowledgments}
We thank Harald Pfeiffer and Sanjay Reddy
for helpful discussions.
This work was supported in part by the Sherman
Fairchild Foundation and NSF Grants No. PHY-1404569 and No. AST-1333520 at
Caltech, NSF Grant No. AST-1333142 at Syracuse University,
the Sherman Fairchild Foundation and 
NSF Grants No. PHY-1306125 and No. AST-1333129 at Cornell University
and by NASA through
Einstein Postdoctoral Fellowship Grant No. PF4-150122 awarded by the
Chandra X-ray Center, which is operated by the Smithsonian Astrophysical
Observatory for NASA under Contract No. NAS8-03060.
Computations were performed on the Zwicky cluster at Caltech,
which is supported by the Sherman Fairchild Foundation and by NSF
Grant No. PHY-0960291; on the NSF XSEDE network under Grant No. TG-PHY990007N; 
on the NSF/NCSA Blue Waters at the University of Illinois with allocation
jr6 under NSF PRAC Grant No. ACI-1440083;
and on the GPC supercomputer at the SciNet HPC
Consortium~\cite{scinet}; SciNet is funded by the Canada Foundation
for Innovation (CFI) under the auspices of Compute Canada; the
Government of Ontario; Ontario Research Fund (ORF)--Research
Excellence; and the University of Toronto.

%

\end{document}